# Dzyaloshinskii–Moriya nature of ferroelectric ordering in magnetoelectric $Gd_{1-x}Y_xMnO_3$ system

R. Vilarinho[1], A. Almeida[1], J. M. Machado da Silva[1], J. B. Oliveira[1], M. A. Sá[1], P. B. Tavares[2],

J. Agostinho Moreira[1]

[1] IFIMUP and IN-Institute of Nanoscience and Nanotechnology, Departamento de Física e Astronomia da Faculdade de Ciências, Universidade do Porto, Rua do Campo Alegre, 687, 4169-007 Porto, Portugal.

[2] Centro de Química - Vila Real, Departamento de Química. Universidade de Trás-os-Montes e Alto Douro, 5000-801 Vila Real, Portugal.

**ABSTRACT**

This work reports on magnetic, dielectric, thermodynamic and magnetoelectric properties of $Gd_{1-x}Y_xMnO_3$, $0 \leq x \leq 0.4$, with emphasis on the (x, T) phase diagram, towards unraveling the role of the driving mechanisms in stabilizing both magnetic and ferroelectric orderings. The (x, T) phase diagram reflects the effect of lattice distortions induced by the substitution of $Gd^{3+}$ ion by smaller $Y^{3+}$ ions, which gradually unbalances the antiferromagnetic against the ferromagnetic exchange interactions, enabling the emergence of ferroelectricity for higher concentrations of yttrium. For $x \leq 0.1$, the paramagnetic phase is followed by a presumably incommensurate collinear antiferromagnetic phase, then a weak ferromagnetic canted A-type antiferromagnetic ordering is established at lower temperatures. For $0.2 \leq x \leq 0.4$, a different phase sequence is obtained. The canted A-type antiferromagnetic arrangement is no more stable, and instead a pure antiferromagnetic ordering is stabilized below $T_{lock} \approx 14 - 17$ K, with an improper ferroelectric character. From these results, a cycloid modulated spin arrangement at low temperatures is proposed, accordingly to the inverse Dzyaloshinskii–Moriya model. Anomalous temperature dependence of the dipolar relaxation energy and magnetization evidence for structural and magnetic changes occurring at T* ≈ 22 - 28 K for $0.1 \leq x \leq 0.4$.

1. Introduction

Magnetoelectric materials are a very interesting class of materials under study. The emergence of ferroelectricity, coupled with the magnetic order, has been viewed as an important issue for fundamental research, as well as, a potential key for technological applications.[1,2]

A huge effort has been done in order to improve the magnetoelectric coupling, namely, at room temperature. A well-known example is the case of $BiFeO_3$, which exhibits ferroelectricity and antiferromagnetic order at room temperature, with an electric polarization along the [111] direction, as large as 100 µC/cm$^2$.[3] Nevertheless, $BiFeO_3$ has large electric conductivity, hindering its use in applications such as barriers and tunnel junctions for spintronics. Substituting the $Bi^{3+}$ by the $La^{3+}$ significantly reduces the electric conductivity, though decreasing significantly the electric polarization.[4,5]

Another class of materials exhibiting magnetoelectricity consists on the rare-earth manganites, with general formula $R$MnO$_3$, with $R$ a rare-earth ion. Among them, the orthorhombic distorted perovskites with $R$ = Eu to Dy exhibit magnetically-induced ferroelectricity at low temperatures.[6] It has been proposed for these compounds that the ferroelectricity is originated from a variety of modulated magnetic structures and can be explained in terms of the model developed by M. Mochizuki and N. Furukawa.[7] The model includes the competitive superexchange interactions, single-ion anisotropy, Dzyaloshinskii–Moriya interaction and the Peierls-type spin-phonon coupling.[7] The competitive superexchange interactions involve the ferromagnetic interaction $J_{ac}$ (P*nma* notation) and the antiferromagnetic interaction $J_2$.[7] While the $J_{ac}$ is weakly dependent on the Mn-O-Mn bond angle, $J_2$ increases with the increase of the orthorhombic distortion, which favors the overlapping of the $Mn^{3+}$ 3d and oxygen 2p electronic orbitals.[7] The balance of the competition between the antiferromagnetic and the ferromagnetic interactions stabilizes the cycloidal-modulated magnetic structure, while the single-ion anisotropy and the Dzyaloshinskii–Moriya interactions define the orientation of the cycloidal plane (*ac* or *ab*), which can be changed by external parameters.[7] According to this model, the shift of the oxygen atom positions, which is dependent on the Peierls-type spin-phonon coupling, modifies the Mn-O-Mn bond angle and, consequently, the $J_2$ superexchange integral, changing the balance of the competition between $J_{ac}$ and $J_2$.[7]

In non-substituted rare-earth manganites, the change of the Mn-O-Mn bonds angle can be achieved by changing temperature, hydrostatic pressure, magnetic field or strain. However, the change of the orthorhombic distortion can be made possible by controlled substitution of the rare-earth ion by smaller ions, such A = $Y^{3+}$ or $Lu^{3+}$.[8–10] This is the case of the solid solutions Eu$_{1-}$ $_x$A$_x$MnO$_3$, which have been extensively studied, both theoretically and experimentally.[11–13] In these systems, the magnetic moments stem only from the Mn$^{3+}$ ions, and both crystal and magnetic structures are strongly dependent on the Y-concentration.[14] As a consequence, these systems exhibit rather interesting (x, T) and (B, T) phase diagrams, which have been explained in the framework of the Mochizuki and Furukawa model.[7,15] The great advantage of these systems is that by increasing Y-concentration only the effect of geometrical mechanisms and thus J$_2$ are expected to influence the phase diagram, enabling to determine their role in defining the nature and number of different phases, including the existence of ferroelectricity in modulated magnetic structures.

Among the rare-earth manganites, GdMnO$_3$ is one of the most outstanding compounds. GdMnO$_3$ undergoes a phase transition into a collinear-sinusoidal incommensurate antiferromagnetic phase at T$_N$ = 42 K, with a spin modulation wave vector (k$_{Mn}$, 1 , 0), if a P*nma* space group is used.[16,17] This modulation corresponds to a collinear arrangement of the Mn$^{3+}$ spins, which is a consequence of competing ferromagnetic and antiferromagnetic exchange interactions between neighbour and next-neighbour Mn$^{3+}$ spins. GdMnO$_3$ undergoes a further magnetic transition into an A-type antiferromagnetic order at T$_{cAFM}$ around 23K.[16,17] The observed weak ferromagnetism has suggested a canted A-type antiferromagnetic ordering of the Mn$^{3+}$ spins along the *b*-axis for T < T$_{cAFM}$.[16,18] The ground state of GdMnO$_3$ in the absence of a magnetic field is not ferroelectric.[19–22] However, by applying a rather low magnetic field (around 10$^4$ Oe) parallel to the *a*-axis, a ferroelectric order is induced along the *c*-axis.[19] Contradictory results have been reported for the ferroelectric polarization. Kuwahara *et al* find a finite polarization below 13 K, while Kimura *et al* observe a ferroelectric order only between 5 K and 8 K, at low magnetic fields.[16,20] The applied magnetic field seems to stabilize a commensurate magnetic structure below 15 K, with a modulation vector δ = 1/4.[17]

Some years ago, Ivanov *et al* have published an experimental work on the orthorhombic Gd$_{1-x}$Y$_x$MnO$_3$, with 0 ≤ x ≤ 0.2.[8] The compounds belonging to this system crystalize in the space group P*nma*, and the unit cell contains four formula units.[8] According to Ref. 8, the collinear-sinusoidal incommensurate antiferromagnetic phase transition undergoes at T$_N$ = 41 K. The most notable result is the suppression of the canted A-type antiferromagnetic ordering, observed in GdMnO$_3$, and the stabilization of a commensurate and, simultaneously, ferroelectric phase below T$_{lock}$ = 20 K, for x ≥ 0.05.[8] A magnetic field of about 20 kOe applied along the *c*-axis, suppresses the commensurate and ferroelectric phase, and induces the transition to the canted antiferromagnetic state.[8] This result is somehow unexpected as the increase of the orthorhombic distortion, induced by the reduction of the A-site size, favor the stabilization of the suitable modulated magnetic structures towards ferroelectricity with an applied magnetic field.[6] The anomalous behavior of the electric permittivity as a function of temperature along with the magnetic field dependence of the electric polarization points out for a magnetoelectric coupling in this system.[8] To the best of our knowledge, no other work has been reported in $Gd_{1-x}Y_xMnO_3$ so far.

The $Gd_{1-x}Y_xMnO_3$ system seems to be more complex than the $Eu_{1-x}A_xMnO_3$, A = $Y^{3+}$ or $Lu^{3+}$, as due to the presence of $Gd^{3+}$, its magnetic properties change differently with the degree of substitution. However, a detailed study of the physical properties in this system is still missing in order to draw a (x, T) phase diagram and to figure out the magnetoelectric properties in $Gd_{1-x}Y_xMnO_3$, with 0 ≤ x ≤ 0.4. In this work, we present a systematic study of the lattice parameters at room conditions, and magnetic, dielectric, ferroelectric, and thermodynamic properties of this system at low temperatures, in order to draw its (*x*, T) phase diagram. A detailed discussion about the underlying mechanisms will be undertaken by analyzing the experimental results in the scope of the mentioned theoretical models.

## 2. Experimental details

High quality $Gd_{1-x}Y_xMnO_3$ ceramics, with x = 0.0, 0.1, 0.2, 0.3 and 0.4, were processed through the urea sol-gel combustion method, sintered at 1623 K for 60 to 90 hours, and then quenched to room temperature. The rapid cooling is known to be efficient to guarantee the oxygen stoichiometry of the samples. The samples were characterized in terms of chemical, morphological and microstructure by using powder X-ray diffractometry, scanning electron microscopy, and X-ray photoemission spectroscopy techniques. Details of sample processing are available elsewhere.[23] The x-ray powder diffraction spectra of the $Gd_{1-x}Y_xMnO_3$ were recorded at room temperature using the X'Perto Pro PANalytical diffractometer, in the Bragg-Bentano geometry. The measurements were performed using the $K_{\alpha 1}$ and $K_{\alpha 2}$ doublet emitted by the Cu cathode, with wavenumbers 1.540598 Å and 1.544426 Å, respectively. The diffractometer uses an X'Celerator detector, with a Ni filter to minimize the $K_\beta$ radiation, and a secondary monocromator. The spectra were measured, in the 10° to the 70° 2θ range, with a step of 0.017° and an acquisition time of 100 s.step$^{-1}$. The calibration and alignment of the diffractometer were made by polycrystalline silica as external standard. The powder diffraction data were analyzed by Le Bail refinements using the FullProf software.[24] The refined parameters were the lattice, the pseudo-Voigt profile and the device calibrations. The XRD spectra background was fitted by linear interpolation between a set of manually chosen points for each diffractogram. The heat capacity was measured in an ARS Cryocooler, between 10 K and 300 K, in a quasi-adiabatic fashion by means of an impulse heating technique. Low-field dc induced specific magnetization measurements were carried out using commercial superconducting quantum interference SQUID magnetometer in the temperature range 5 K to 300 K, with a resolution better than $5 \times 10^{-7}$ emu. Rectangular parallelepipedic shape samples were prepared from the ceramic pellet, and gold electrodes were deposited using the evaporation method. The same samples were used to measure the complex electric permittivity and the thermally stimulated depolarizing currents. The complex electric permittivity was measured with an HP4284A impedance analyzer. The measurements were performed under an ac electric field of amplitude around 1 V.cm$^{-1}$ in the frequency range from 1 kHz to 1 MHz. The thermally stimulated depolarizing currents were measured as a function of temperature, with a standard short-circuit method, using a Keithley 617 electrometer, with a resolution of 0.1 pA, in a heating run, keeping a fixed temperature rate of 5 K/min, after cooling the sample under a poling electric field. The sample temperature was measured with an accuracy better than 0.1 K. The same experimental setup was integrated in a SQUID insert, to measure the thermally stimulated depolarizing current down to 5 K, under magnetic field, up to 5 T, perpendicularly applied to the poling electric field.

## 3. Experimental results

### a. Lattice parameters and distortions at room conditions

Figure 1 shows the experimental x-ray powder diffraction pattern of the $Gd_{1-x}Y_xMnO_3$, with x = 0.0 and 0.4, as representative members of the system, recorded at room conditions, in the 10° to the 70° in 2θ range. The x-ray diffraction pattern of GdMnO$_3$ exhibits the typical spectral profile observed for orthorhombic rare-earth manganites, with crystal structure described by the P*nma* space group.[25,26] For the compounds with x = 0.1 up to 0.4, the x-ray diffraction patterns are quite similar to the GdMnO$_3$ ones, except from the shift of the diffraction peaks toward higher 2θ values as the Y-content increases. The peak shifts are consequence of the volume reduction, due to the substitution of Gd$^{3+}$ ion by the smaller Y$^{3+}$ one, considered as a compressive hydrostatic pressure. The x-ray patterns for the range of compositions with 0 ≤ x ≤ 0.4, are consistent with the P*nma* space group. The unit cell of all studied compounds has four formula units.

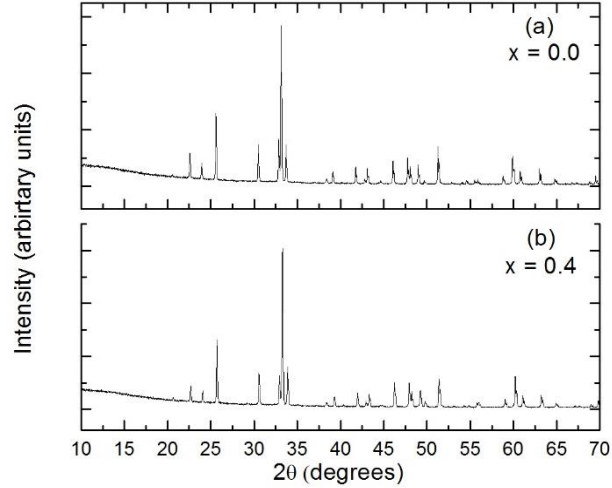

Figure 1. X-ray powder diffraction pattern of $Gd_{1-x}Y_xMnO_3$, with (a) x = 0.0 and (b) x = 0.4, recorded at room conditions.

The distortions induced by the Y-incorporation can be studied by analyzing the x-dependence of the lattice parameters. Perovskites with a tolerance factor smaller than unity, as it is the case of the $Gd_{1-x}Y_xMnO_3$, often exhibit octahedral tilting. In the presence of in-phase octahedra tilting around the [010] axis, a relatively large expansion of the *a*-axis and relatively small change in the length of the *c*-axis are observed. Figure 2 shows the pseudocubic lattice parameters as a function of the Y-content, defined as $a_{pc} = a/\sqrt{2}$, $b_{pc} = b/2$, $c_{pc} = c/\sqrt{2}$, where *a*, *b* and *c* are the lattice parameters of the P*nma* structure. The pseudocubic lattice parameters well satisfy the relation $a_{pc} > c_{pc} > b_{pc}$, which has been typically found in perovskites presenting both octahedra tilting and Jahn-Teller distortion.[27] As x increases, a linear decrease of the lattice parameters is evident. The slope of the $a_{pc}(x)$ relation is three times smaller than the $c_{pc}(x)$, meaning that as x increases, the difference between *a* and *c* increases, which reflects an increasing of the tilting of the $MnO_6$ octahedra around the *c*-axis, and implies the decrease of the Mn-O-Mn bond angle within the *ac*-plane. The decrease of the Mn-O-Mn bond angle unbalances the competition between the ferromagnetic $J_{ab}$ and antiferromagnetic $J_2$ exchange interactions, favoring the antiferromagnetic against the ferromagnetic interactions, as it is well established in the model of Mochizuki and N. Furukawa.[15] This mechanism yields alterations of the spin arrangement at low temperatures and, hence, tunes the corresponding magnetoelectric coupling and ferroelectric properties, as it will be discussed later.

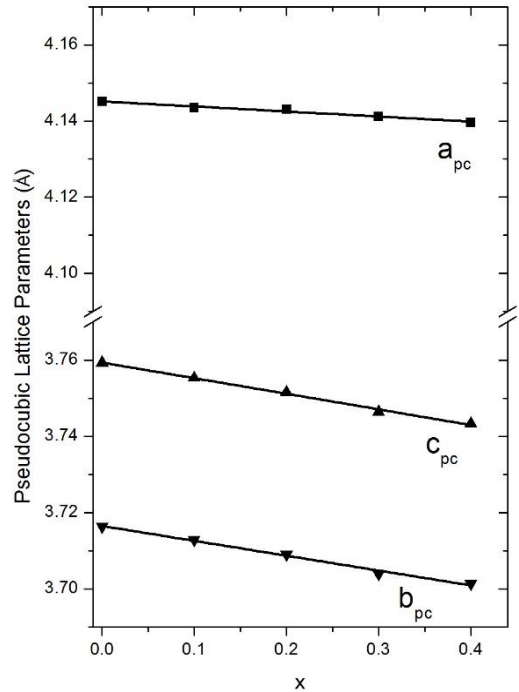

Figure 2. Pseudo-cubic lattice parameters as a function of x, obtained at room conditions. The solid lines were determined by fitting a linear equation to the experimental results.

b. Specific heat

Figure 3 shows the temperature dependence of the specific heat divided by temperature, for the composition x = 0.0 and 0.4, in the 10 K to 160 K temperature range, chosen as representative results. A well-defined lambda-like anomaly of the C/T curve is observed at around 42 K, being practically independent on Y-amount. For GdMnO$_3$, this anomaly is associated with the ordering of the Mn$^{3+}$ spins from the paramagnetic to the collinear-sinusoidal incommensurate antiferromagnetic phase, as reported in the literature.[21] This magnetic phase transition has been also observed for EuMnO$_3$, TbMnO$_3$, DyMnO$_3$ and Eu$_{1-x}$A$_x$MnO$_3$, with A = Y$^{3+}$ or Lu$^{3+}$ and x from 0 up to 0.5.[6,10,16] The critical temperature is not strongly dependent on the A-site cation size, taking values between 42 K and 45 K.[6] As in Gd$_{1-x}$Y$_x$MnO$_3$, x= 0.1 to 0.4, the temperature where the lambda-like anomaly occurs in the C/T curve is nearly independent on Y-amount, and taking into account the same assignment on the doped EuMnO$_3$ system, we assume that for these compositions a similar paramagnetic to a collinear-sinusoidal incommensurate antiferromagnetic phase transition also occurs at the Néel temperature T$_N$ = 42 K. The continuous lines shown in the Figure 3 were determined, according to the Debye model, from the best fit of the lattice contribution to the specific heat between 100 K and 300 K, to the equation:[28]

$$C_D = 9N_h k_B \left(\frac{T}{\theta_D^h}\right)^3 \int_0^{\frac{\theta_D^h}{T}} \frac{x^3}{e^x-1} dx + 9N_l k_B \left(\frac{T}{\theta_D^l}\right)^3 \int_0^{\frac{\theta_D^l}{T}} \frac{x^3}{e^x-1} dx \qquad (1)$$

where $N_h$ and $N_l$ stand for the number of heavy and light atoms, respectively. In our case, $N_h = 2$ and $N_l = 3$. Furthermore, $k_B$ is the Boltzmann constant, and $\theta_D^h$ and $\theta_D^l$ are fitting parameters; $\theta_D^h$ and $\theta_D^l$ are the Debye temperatures for heavy and light atoms, respectively. Table 1 shows the values of the Debye temperatures for heavy and light atoms calculated from the fitting procedure for each composition. While $\theta_D^h$ takes values of the order of 300 K, $\theta_D^l$ is of the order of 800 K. The values obtained in the in $Gd_{1-x}Y_xMnO_3$, with $0 \leq x \leq 0.4$, are in good agreement with those obtained in the $Eu_{1-x}Y_xMnO_3$ system, with $0 \leq x \leq 0.5$.[29] The resemblance of the Debye temperatures for both systems points out for the similarity of the crystalline orthorhombic structure of both $Gd_{1-x}Y_xMnO_3$ and $Eu_{1-x}Y_xMnO_3$ systems.

The area between the experimental curve and the extrapolated Debye behavior below 100 K refers to the magnetic contribution to the specific heat, and it is associated with the magnetic phase transitions taking place at low temperatures. The insets of Figure 3 show the magnetic contribution of the specific heat as a function of temperature, $C_{mag}(T)$, for the compositions with x = 0 and 0.4. The entropy variation associated with the magnetic phase transitions at low temperatures is given by the integral of the magnetic contribution to the specific heat:[30]

$$\Delta S = \int_{10}^{T_M} \frac{C_{mag}}{T} dT \qquad (2)$$

where $T_M$ is the maximum registered temperature. The value obtained for the entropy variation of $GdMnO_3$ is 14.8 J.K$^{-1}$.mol$^{-1}$, while for the remaining compositions is 13.7 J.K$^{-1}$.mol$^{-1}$. The theoretical entropy variation associated with the magnetic transition as the $Mn^{3+}$ spins order can be calculated by:[30]

$$\Delta S = N_A k_B \ln(2s + 1) \qquad (3)$$

where $\Delta S$ is the maximum entropy variation obtained from Eq. (2), $N_A$ is the Avogadro number and is the total spin quantum number. $Mn^{3+}$ ion is known to have = 2, in its high spin configuration,[31] thus the expected entropy variation associated with the $Mn^{3+}$ spin ordering is 13.4 J.K$^{-1}$.mol$^{-1}$.

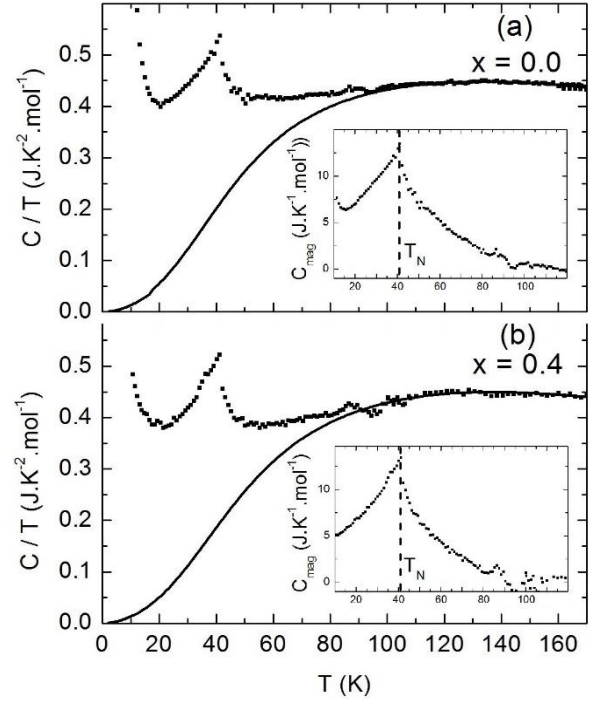

Figure 3. Temperature dependency of the specific heat divided by temperature (full symbols), for the composition (a) x = 0.0 and (b) x = 0.4. The solid line was obtained from the best fit of Eq. (1) to the experimental data recorded above 100 K. The insets show the temperature dependence of the magnetic contribution to the specific heat. The vertical dashed line marks $T_N$ = 42 K.

| Composition | $\theta_D^h$ (K) | $\theta_D^l$ (K) |
|---|---|---|
| x = 0.0 | 314 | 771 |
| x = 0.1 | 326 | 750 |
| x = 0.2 | 319 | 765 |
| x = 0.3 | 315 | 818 |
| x = 0.4 | 326 | 754 |

Table 1. Debye temperatures for heavy and light atoms, obtained from the best fit of Eq. (1) to experimental data, recorded above 100 K.

The experimental values obtained for $GdMnO_3$ and the compositions $0.1 \leq x \leq 0.4$ are about 10% and 2% larger than the theoretical expected values, respectively. The former value evidences a non-negligible contribution of the $Gd^{3+}$ spins to the entropy variation, well above the Néel temperature $T_N^{Gd}$ = 6 K,[16] which is associated with the $Gd^{3+}$ spin ordering. This assumption can also be ascertained by the difference between the curves shown in the insets of Figure 3 for T < 12 K. The quite similar values found for the entropy variation for the other compositions reflect the effect of the structural A-site disorder, which weakens the precursor magnetic interactions associated with the ordering of the $Gd^{3+}$ ion spins at $T_N^{Gd}$. However, it cannot be interpreted as the $Gd^{3+}$ spins do not contribute to the magnetic properties of the system above 12 K, namely under an applied magnetic field, as it can be seen in the next section.

### c. Magnetic properties

Figure 4 shows the temperature dependence of the molar magnetization for x = 0.0, 0.1, 0.2 and 0.4, measured in zero field-cooling (ZFC) and field-cooling (FC) conditions, under a DC applied magnetic field of 40 Oe. Let us first address the paramagnetic phase. Inset of Figure 4(d) shows the temperature dependence of H/M, similar to the inverse of the molar magnetic susceptibility, in the paramagnetic phase. From the linear behavior of H/M(T) above 70 K, we can conclude that all compounds closely follow the Curie-Weiss law:

$$\chi(T) = \frac{C}{T - \theta_c} \qquad (4)$$

where C is the Curie constant and  is the Curie-Weiss temperature. Fitting Eq. (4) to the H/M(T) data above 80 K, the Curie-Weiss temperature $\theta_c$ and the effective paramagnetic moment $\mu_{eff}$ were determined. Table 2 shows the values of the Curie-Weiss temperature, and both the experimental and theoretical values of the effective paramagnetic moment for the studied compositions. The Curie-Weiss temperature $\Theta_c$ takes almost the same value, close to -40 K, indicating the importance of the antiferromagnetic correlations in the paramagnetic phase. The experimental value for $\mu_{eff}$ decreases for increasing Y-concentration, as a consequence of the substitution of the $Gd^{3+}$ ion by the non-magnetic $Y^{3+}$ one. The obtained values are in good agreement with those ones (see Table 2) calculated by taking into account the experimental value of the magnetic moment of $Mn^{3+}$ ( = 4.80 $\mu_B$) and the tabulated one for $Gd^{3+}$,[32] according to equation:

$$\mu_{eff}^2 = (1-x)\mu_{Gd^{3+}}^2 + \mu_{Mn^{3+}}^2 \qquad (5)$$

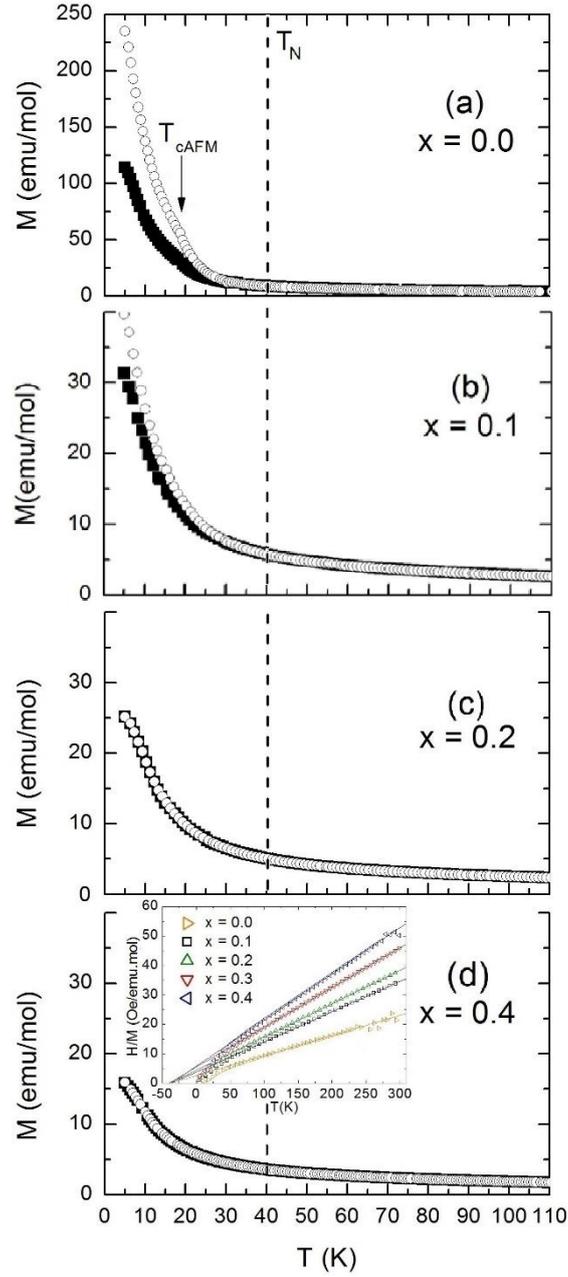

Figure 4. Temperature dependence of the molar magnetization measured in zero-field and field cooling conditions, under a 40 Oe magnetic field, for the $Gd_{1-x}Y_xMnO_3$ with (a) x = 0.0, (b) 0.1, (c) 0.2 and (d) 0.4. Inset shows the H/M ratio, for x = 0.0 to 0.4. The solid line was calculated from the best fit of *Eq. (5)* to the experimental data, above 80 K.

| Composition | $\Theta_c$ (K) | $\mu_{eff}$ ($\mu_B$) | $\mu_{eff,theo}$ ($\mu_B$) |
|---|---|---|---|
| x = 0.0 | -38 | 9.20 | 9.33 |
| x = 0.1 | -37 | 8.82 | 8.99 |
| x = 0.2 | -40 | 8.41 | 8.63 |
| x = 0.3 | -40 | 7.69 | 8.25 |
| x = 0.4 | -38 | 7.16 | 7.86 |

Table 2 - Curie-Weiss temperature and the experimental and calculated effective paramagnetic moment as a function of the Y-concentration.

In the following, we will discuss the low temperature magnetic phase transitions. First of all, we stress that the increase of molar magnetic response at low temperatures is a consequence of the $Gd^{3+}$ spin ordering, which takes place at $T_N^{Gd}$ = 6 K.[33] This interpretation is consistent with the decrease of the magnitude of the molar magnetization as the Y-concentration increases. In fact, $Gd^{3+}$ is known to large contribute to the magnetic response well above its critical temperature. Moreover, the magnetic contribution coming from the $Mn^{3+}$ spin system is superimposed and affects the magnitude of the magnetic response measured in ZFC and FC conditions. Actually, the magnetic contribution of the $Mn^{3+}$ spins to the magnetic response is apparent in the difference between the magnitude of the ZFC and FC magnetic response curves, as it will be discussed in the following.

$GdMnO_3$ is characterized by a much larger magnitude of the magnetic response and a difference between the FC and ZFC curves (see Figure 4(a)). The increase of the magnetic response in FC conditions reveals a weak ferromagnetic phase in the $GdMnO_3$ compound, stable below 20 K.[18] The difference in the magnetic response emerges from the $Mn^{3+}$ spin canting, which is characteristic of the weak ferromagnetism associated with the canted A-type antiferromagnetic phase, stable below $T_{cAFM}$ = 20 K. Besides, it is similar to what was reported for $RMnO_3$, with $R$ = Eu to Dy.[6] Moreover, a small but significant anomaly is detected in the M(T) curves around 20 K. This anomaly will be addressed later.

The magnetic response of the compositions with 0.2 ≤ x ≤ 0.4, shows a superposition of the ZFC and FC curves, and no enhancement of the magnetic response is observed in the presence of a DC magnetic field during the cooling run, as it can be observed from Figs. 4(c) and (d). The absence of an increase in the magnetic response under FC conditions points to a reinforcement of the antiferromagnetic interactions between the $Mn^{3+}$ spin system against the ferromagnetic ones. This result corroborates the decrease of the Mn-O-Mn bond angle, as it was inferred from the structural data referred to above. The absence of a typical antiferromagnetic M(T) curve, which is characterized by a gradually decrease of the magnetic response as temperature decreases below the Néel temperature, is a consequence of the magnetic response of the system to the $Gd^{3+}$ spin ordering at low temperature, that superimpose to the response of the $Mn^{3+}$ spins.

The understanding of the M(T) curves for the case of x = 0.1 is more complex. The difference between the FC and ZFC curves is non-vanishing, but it is far less than for $GdMnO_3$. Moreover, the magnitude of the magnetic response is comparable with the one obtained for the compositions with higher x. The small difference between the FC and ZFC curves points to a weak ferromagnetic character of this composition. However, due to the decrease of the Mn-O-Mn bond angle, already ascertained from the x-ray data, we expect an increase of the antiferromagnetic character of $Gd_{0.9}Y_{0.1}MnO_3$ relatively to $GdMnO_3$. Figure 5 shows the magnetization as a function of the applied magnetic field, measured in increasing and decreasing magnetic strength at 5 K, for x = 0.0 and 0.1.

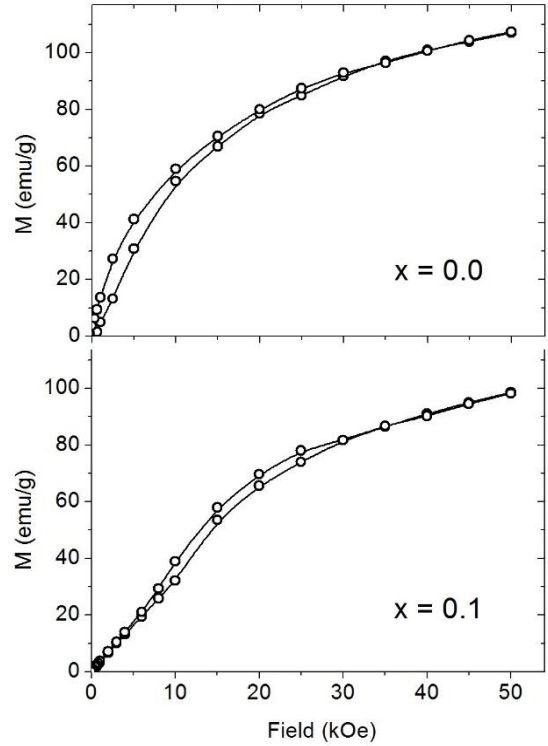

Figure 5. Magnetization as a function of the applied magnetic field strength, measured in increasing and decreasing magnetic field, for x = 0.0 and 0.1 at fixed temperature T = 5 K.

For $GdMnO_3$, a well-defined ferromagnetic hysteresis loop is observed. This result is a strong evidence of the weak ferromagnetic character of $GdMnO_3$, due to spin canting below $T_{cAFM}$ = 20 K.[18] Moreover, at this temperature, a small but clear anomaly in both ZFC and FC curves is observed (see Figure 4(a)). This temperature is in good agreement with the critical temperature reported for $GdMnO_3$ associated with the transition from the collinear-sinusoidal incommensurate antiferromagnetic to the canted A-type antiferromagnetic phase.[6] For the composition with x = 0.1 instead of a typical ferromagnetic hysteresis loop, a linear M(H) relation in the low-strength magnetic field range (H < 5 kOe) is observed at 5 K. A clear change of the M(H) relation is detected for magnetic fields larger than 10 kOe, where a hysteresis emerges, characteristic of an antiferromagnetic response. The critical magnetic field is about 5 kOe. The aforementioned result is in good agreement with the M(H) behavior published by Ivanov *et al*[8], and corroborates the increase of the antiferromagnetic character of the $Gd_{0.9}Y_{0.1}MnO_3$. Contrarily to the $GdMnO_3$ case, the weak ferromagnetic response of $Gd_{0.9}Y_{0.1}MnO_3$, ascertained from the difference between the FC and ZFC M(T) curves, could not be observed in the M(H) curves due to reduced value of the ferromagnetic component.

### d. Dielectric properties

The left panel of Figure 6 exhibits the real (Ɛ') and the right panel the imaginary (Ɛ'') parts of the complex electric permittivity for x = 0.0, 0.1, 0.2, 0.3 and 0.4, as a function of temperature, measured at 50kHz, 100kHz and 1MHz, respectively. Measurements were both done in heating and cooling runs, and no thermal hysteresis was found. It is worthwhile to note that for these materials the Ɛ'(T) curves obtained in this work resemble the Ɛ'(T) curves obtained along the *a*-direction of single crystals.[8] Two types of anomalies are evidenced in the electric permittivity curves. The first set of anomalies, we will consider, consists of an over-spread step in Ɛ'(T) and a broad peak in Ɛ''(T), both strongly dependent on the measurement frequency, and can be observed in the temperature dependence of the electric susceptibility of both sets. From the selected frequencies displayed in Figure 6, it is clear that the step of Ɛ'(T) and the maximum of the broad peak in Ɛ''(T) shifts to higher temperatures as frequency increases. This behavior is associated with a thermal-driven dielectric dipolar relaxation and, like in other rare-earth manganites, it is not assignable to any critical phenomena, suggesting a non-cooperative mechanism driving this relaxation.[34] The study of the dielectric relaxation was carried out varying the temperature at constant frequency. In this case, the relaxation behavior is better analyzed by fitting the Debye model to the Ɛ''(T) curve, according to equation:[35]

$$\varepsilon''(\omega,T) = \frac{\frac{1}{2}[\varepsilon(0)-\varepsilon(\infty)]}{cosh\left[\frac{U}{k_B}\left(\frac{1}{T}-\frac{1}{T_m}\right)\right]} \quad (6)$$

where $T_M$ is the temperature of the anomaly maximum in Ɛ''(T), U is the activation energy and stand for the lower and upper frequency limits of the electric permittivity, respectively.

Figure 7(a) shows an example of the fitting of Eq. (6) to the Ɛ''(T) curve of $Gd_{0.9}Y_{0.1}MnO_3$ measured at 500 kHz, in the 48 K to 70 K range. The frequency-dependent anomaly of Ɛ''(T) is well described by Eq. (6), pointing for a monodispersive character of the relaxation mechanism. This assumption is corroborated by the Cole-Cole plot, shown in the inset of Figure 7(a), where we can see that the experimental (Ɛ', Ɛ'') points lie on a semicircle with its center on the Ɛ' axis.

The deviation of the Ɛ''(T) curve from the extrapolated Debye behavior, described by Eq. (6), at high temperatures comes from the conduction mechanism present in the samples. Moreover, the magnetic phase transition occurring at $T_N$ is well marked by the deviation of the Ɛ''(T) curve from the Debye behavior just below 45 K. A small anomaly is also detected in the Ɛ''(T) curve at around T* = 20 K. This anomaly reveals a further critical temperature that will be discussed in more detail later on.

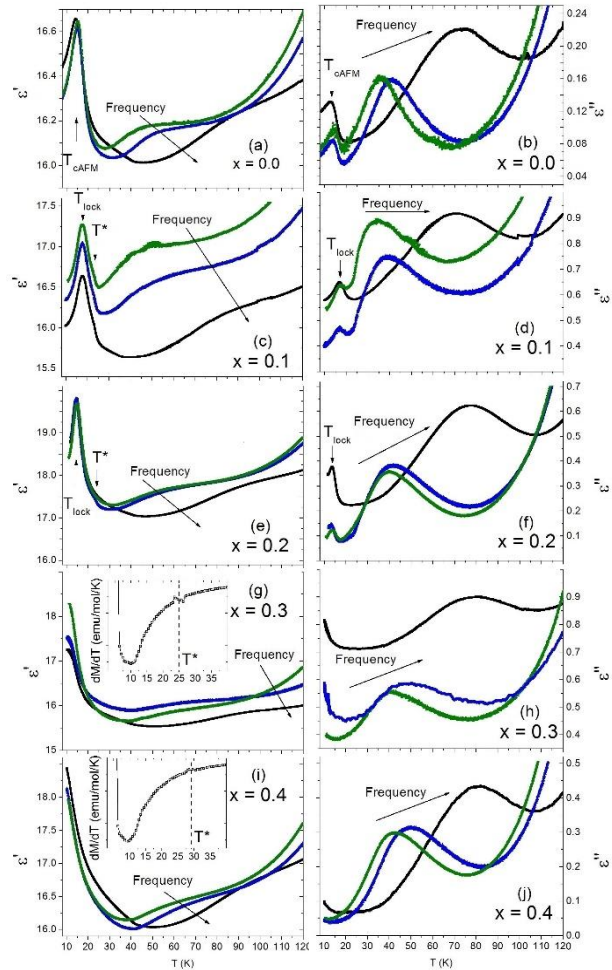

Figure 6. In the left panel, the real Ɛ' and in the right panel, the imaginary Ɛ'' parts of the electric permittivity of x = 0.0 to 0.4, measured at the fixed frequencies of 50kHz, 100kHz and 1MHz, in heating run. The insets of Figures 4(g) and 4(i) show the temperature dependence of the first derivative of the magnetic response for x = 0.3 and 0.4.

From the aforementioned fit procedure, $T_m$ and U were determined for each frequency in each compound. However, a better determination of U can be achieved if the relationship between relaxation frequency and temperature is analyzed. Figures 7(b) and (c) shows the logarithm of the relaxation frequency as a function of $T^{-1}$, for x = 0.0 and 0.4, chosen as representative compositions of the two sets. As it can be seen, linear relationships are obtained in well-defined temperature ranges, showing an Arrhenius behavior of the relaxation time. The change of the slope of the linear relations reflects the modification of the activation energy, which takes values from 10 meV to 30 meV, as it is common for dipolar relaxation mechanisms observed in GdMnO$_3$ and Eu$_{1-x}$Y$_x$MnO$_3$.[36,37] For each composition, distinct temperature dependences of the relaxation frequency are observed, but the limits of the corresponding temperature ranges are often hard to define. Nonetheless, the change of slope of the linear relations reflects the occurrence of structural changes at low temperatures, which is reflected in the dipolar activation energy. The first slope change occurs around the Néel temperature $T_N$, in accordance with the anomaly in the C/T(T) and Ɛ''(T) curves, already mentioned. This fact reveals that the magnetic phase transition, taking place at $T_N$, modifies the activation energy of the dipolar relaxation process. For the case of x = 0.0, the results obtained from the complex electric permittivity measurements at low frequencies, below 50 kHz, are noisy, preventing the study of the relaxation process below 35 K. However, for the compositions with x = 0.1 to 0.4, another change of slope is observed close to T*, ranging from 26 to 34 K depending on x. This feature suggests another structural transformation that also modifies the activation energy of the relaxation process, as it will be discussed in the next section. In order to determine the mean value of the activation energy in each temperature range, as well as, the experimental error, we have fitted a linear function to the results, while choosing different number of points. The obtained values for the activation energy following this procedure, differ less than 1 eV. The results obtained for the activation energy and the corresponding error in each temperature range are listed in Table 3. The activation energies for the pure GdMnO$_3$ are consistent with those presented in a previous published work.[36] The activation energy of the relaxation process takes the maximum value in the paramagnetic phase above $T_N$, and decreases as the phase sequence is scanned down to low temperatures. In each phase, the activation energy is constant, as the linear relation of the logarithm of the relaxation frequency against $T^{-1}$ confirms. The activation energy of the dipolar relaxation depends on the structure of the compound. In other rare-earth systems (GdMnO$_3$, Eu$_{1-x}$Y$_x$MnO$_3$, Eu$_{1-x}$Lu$_x$MnO$_3$) the antiferromagnetic phase transition taking place at $T_N$ reveals itself as anomalous temperature behavior of the optical phonon frequencies, measured by Raman scattering.[36–38] This result has been pointed out as a clear evidence for the spin-phonon coupling in those rare-earth manganites.[9,13,39] The change of frequency of the optical phonons are a consequence of structural rearrangements induced by the magnetic order, which change the activation energy. So, the change of activation energy at well-defined temperatures evidenced in this work, which depend on the Y-concentration, points to changes on the energy barriers, due to magnetic spin rearrangements at T*. Although the dipolar relaxation process proved in this work is not associated with any critical phenomena, the magnetically-induced changes in the energy barriers clearly support the existence of a magnetodielectric coupling in these compounds.

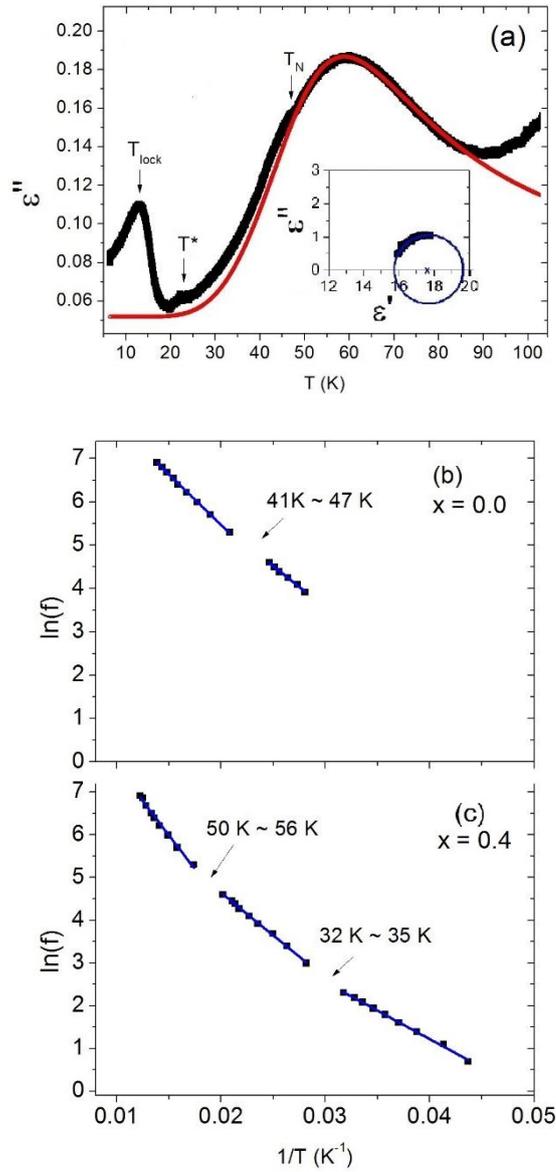

Figure 7. (a) Imaginary part of the electric permittivity of $Gd_{0.9}Y_{0.1}MnO_3$ as a function of temperature, measured at 500 kHz. The solid line was calculated from the best fit of Eq. (6) to the $\varepsilon''(T)$ data, in the 48 K to 70 K temperature range. Inset of Figure (a) shows the Cole-Cole plot, using the experimental data recorded at 30 K. Logarithm of the relaxation frequency as a function of $T^{-1}$ for (b) x = 0.0 and (c) x = 0.4.

|             | U (meV)   |                  |           |
|-------------|-----------|------------------|-----------|
| Composition | T > $T_N$ | $T_N$ > T        |           |
| x = 0.0     | 20 ± 1    | 17 ± 1           |           |
|             | T > $T_N$ | $T_N$ > T > $T^*$ | $T^*$ > T |
| x = 0.1     | 19 ± 1    | 12 ± 1           | 11 ± 1    |
| x = 0.2     | 20 ± 1    | 17 ± 1           | 10 ± 1    |
| x = 0.3     | 31 ± 1    | 17 ± 1           | -         |
| x = 0.4 | 27 ± 1 | 17 ± 1 | 11 ± 1 |

Table 3. Activation energy of the relaxation process as a function of Y-amount, obtained from the linear relations in different temperature ranges.

The anomalies in the $\varepsilon'(T)$ and $\varepsilon''(T)$ curves, located below 42 K, are associated with cooperative phenomena and form the second type of anomalies evidenced in this study. The transition from the paramagnetic to the collinear-sinusoidal incommensurate antiferromagnetic phase is revealed by a clear anomaly in the C/T (T) curve and a tiny anomaly in $\varepsilon''(T)$ at $T_N$ (see Figure 7(a)). Due to the relaxation mechanisms already referred to above, this anomaly is only visible in the $\varepsilon'(T)$ and $\varepsilon''(T)$ curves measured at certain frequencies. Such kind of anomaly is also observed in the $\varepsilon'(T)$ and $\varepsilon''(T)$ curves obtained in other compositions, and the value of $T_N$ is quite independent of x, in good agreement with the values obtained from the analysis of the specific heat curves. Let us return to Figure 6, where the real and imaginary parts of the complex electric permittivity of $Gd_{1-x}Y_xMnO_3$, with x = 0.0, 0.1, 0.2, 0.3 and 0.4, as a function of temperature are presented. In the case of $GdMnO_3$, the peak in both $\varepsilon'(T)$ and $\varepsilon''(T)$ curves at $T_{cAFM}$ = 17 K marks the transition from the collinear-sinusoidal incommensurate antiferromagnetic phase into the canted A-type antiferromagnetic phase, in good agreement with previous reports concerning this transition.[16,37] For the compounds with x = 0.1 and 0.2, a well-defined peak-like anomaly is observed both in $\varepsilon'(T)$ and $\varepsilon''(T)$ curves, which peaks at $T_{cAFM}$ = 18 K and $T_{lock}$ = 14 K respectively. The amplitude of the peak of the real part increases from 0.7 for x = 0.1, to 2.7 for x= 0.2. Both the maximum temperature and amplitude of the $\varepsilon'(T)$ anomaly here reported for the composition with x = 0.1 are in good agreement with the data published elsewhere for single crystals with this composition.[8] This anomaly shifts to lower temperatures for the x = 0.3 and 0.4 compositions, in such a way that the peak maximum is no longer observed in the measuring temperature range.

The $\varepsilon'(T)$ curve for the compositions x = 0.1 and 0.2 also reveals a shoulder-like anomaly occurring at about T* = 22 K. In the $\varepsilon''(T)$ curve for the x = 0.1 compound a tiny anomaly at the same temperature is observed (see Figure 6(a)). No hint of any anomalous behavior could be observed in $\varepsilon'(T)$ and $\varepsilon''(T)$ for the other compositions in the vicinity of 22 K. However, a detailed analysis of the temperature derivative of the magnetization, dM/dT(T), of the compositions with x = 0.3 and 0.4, and depicted in the insets of Figures 6(g) and 6(i), shows a small but clear anomalous temperature behavior at T* = 25 K and 28 K, respectively. This kind of anomaly observed at T* in both electric permittivity and magnetization was not previously reported.

d. Polar properties

The results reported above yield clear evidence that as x increases the Mn-O-Mn bond angle decreases. Regarding the Mochizuki and Furukawa model for rare-earth manganites,[15] the decrease of the Mn-O-Mn bond angle strengths the antiferromagnetic interactions against the ferromagnetic ones. This is clearly ascertained from the magnetic response of the $Gd_{1-x}Y_xMnO_3$ system, where the weak ferromagnetic behavior, observed for the composition x = 0.0, is suppressed with increasing yttrium concentration, turning eventually into a pure antiferromagnetic state. For the concentrations where the antiferromagnetic state is stabilized a ferroelectric polarization is expected to coexist with a cycloidal commensurate modulated spin arrangement, similar to that observed in $TbMnO_3$, $DyMnO_3$ and Y-doped $EuMnO_3$.[6,9,10]

By keeping these main outcomes in mind, we will present and discuss the results regarding the polar properties of the $Gd_{1-x}Y_xMnO_3$ system, measured using the thermally stimulated depolarization currents technique, using applied magnetic fields up to 5 T. For simplicity, we will start by discussing the parent compound, $GdMnO_3$, without any magnetic field. Inset of Figure 8(a) shows the current density as a function of temperature, measured in a heating run, at a temperature rate of 5 K min$^{-1}$, after cooling the $GdMnO_3$ sample with a poling electric field (830 V/cm) below 50 K. A peak in the temperature dependence of the current density is observed, around $T_{cAFM}$ = 17 K, where the $\varepsilon'(T)$ and $\varepsilon''(T)$ curves display anomalous behavior (see Figures 6(a) and 6(b)). However, according to the Dzyaloshinskii-Moriya model, the canted A-type antiferromagnetic phase, established in $GdMnO_3$, does not allow for the stabilization of a ferroelectric phase.[15] Thus, we have to consider this polarization as arising from an induced mechanism, rather than a cooperative phenomenon. Following this assumption, the current peak observed was analyzed by considering the existence of an induced polarization. The thermally stimulated depolarization current peak is described by the equation:[40]

$$J_D(T) = \frac{P_e(T)}{\tau_0} \exp\left(-\frac{U}{k_B T}\right) exp\left(-\frac{1}{q\tau_0} \frac{k_B T^2}{U} e^{-\frac{U}{k_B T}}\right) \quad (7)$$

where $P_e$ is the equilibrium polarization, $\tau_0$ is the relaxation time at infinite temperature, U is activation energy of dipolar orientation and q = dT/dt is the heating rate. The result obtained from fitting equation (7) to the experimental data obtained at zero magnetic field is shown in Figure 8(a). The superposition observed between fitting curve and experimental data apparently corroborates the induced nature of the electric polarization in $GdMnO_3$, as it was expected from the considerations referred to above. The obtained parameters from the fit procedure are $P_e$ = 17.0±0.2 pC/cm$^2$, $\tau_0$ = 0.9±0.2 ms and U = 16.2±0.3 meV. The composition with x = 0.1 also presents a peak on the current density curve, close to $T_{cAFM}$ = 18 K, cooled with a poling electric field of 500 V/cm. Similar analysis was carried, and the obtained fit parameters are $P_e$ = 26.2±0.3 pC/cm$^2$, $\tau_0$ = 2.3±0.5 ms and U = 13.7±0.3 meV. For the two compounds in discussion, the value of the activation energy obtained in this analysis is similar to those measured for dipolar relaxations in $Eu_{1-x}Y_xMnO_3$.[37] After subtracting the fitted curve to the experimental data, we can obtain the density current associated with the cooperative phenomenon in the absence of any applied magnetic field. The result of this procedure is shown in Figure 8(a) for $GdMnO_3$ and $Gd_{0.9}Y_{0.1}MnO_3$. As it can be ascertained, no electric polarization is established without applied magnetic field.

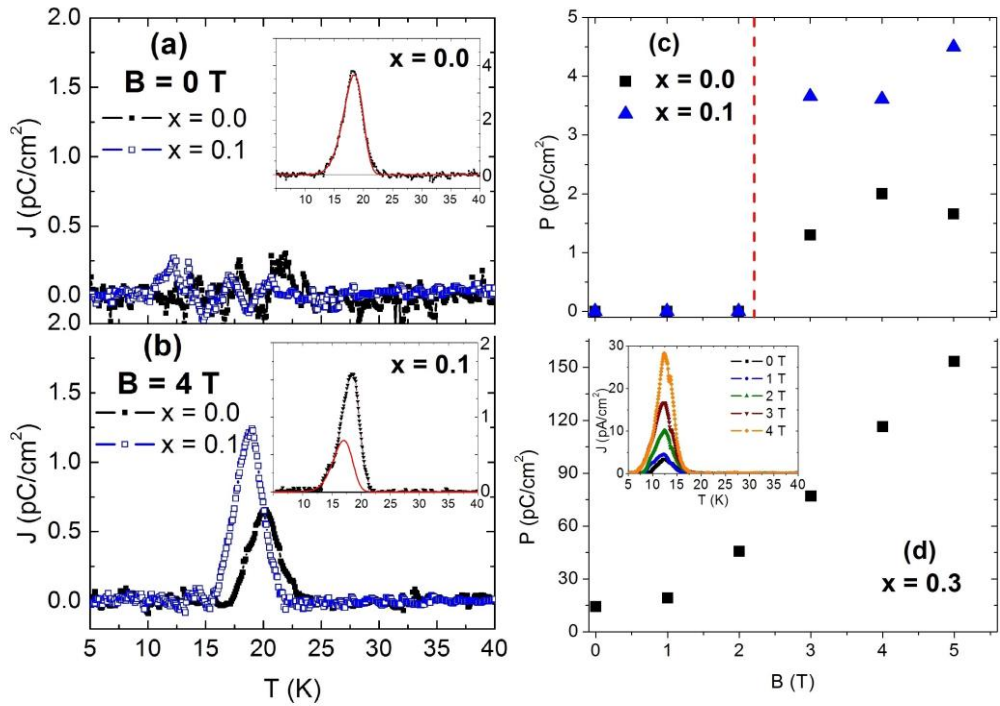

Figure 8. Temperature dependence of the current density for x = 0.0 and x = 0.1, (a) without applied magnetic field and (b) with applied magnetic field of 4 T. Insets of (a) and (b) shows the best fit of Equation (7) to the experimental data (solid line) as described in the text. Electric polarization at 5 K as a function of applied magnetic field for x = 0.0 and 0.1 (c); and for x = 0.3 (d). Dashed vertical line signalizes the critical field for $GdMnO_3$, according to Ref 21. Inset of (d) shows the temperature dependence of the current density for x = 0.3 for several applied magnetic fields.

The same analysis was performed under applied magnetic field, maintaining both the temperature rate and the poling electric field strength. The magnetic field is known to promote the emergence of the ferroelectric phase in several $RMnO_3$.[17,19] Assuming that both the relaxation time and activation energy are weakly dependent on the magnetic field, we have fitted equation (7) by varying only the equilibrium polarization. This procedure will enable us to subtract the induced component of the polarization, and determining the magnetically induced ferroelectric polarization. Inset of Figure 8(b) shows an example of the fitting procedure results for the composition x = 0.1, where it is observed that both induced and ferroelectric phenomena coexist. Figure 8(b) presents the subtraction of the fitted curve to the experimental data, for the example of x = 0.1, under an applied magnetic field of 4 T. After integrating the density current we have determined the value of the ferroelectric polarization at 5 K as function of magnetic field, and the result is shown in Figure 8(c). A clear emergence of the electric polarization for B > 2 T is ascertained, increasing with increasing magnetic field, as it was also observed in previously published works.[17,19]

For x > 0.1, the antiferromagnetic phase is well established, and thus a ferroelectric phase is expected even in the absence of the magnetic field. In order to ascertain this statement, we present, as a representative example, the results obtained in the composition x = 0.3, after cooled down the sample under a poling electric field of 900 V/cm. In the inset of Figure 8(d) we can see that the area under the density current curve steadily increases as the magnetic field strength increases, which gives clear evidence for the ferroelectric nature of the observed thermally stimulated current density. The magnetic field strength dependence of the spontaneous polarization at 5 K is depicted in Figure 8(d).

The main outcome of this analysis is that for higher concentrations of $Y^{3+}$ the ferroelectric phase is stabilized, even in the absence of an applied magnetic field. Moreover, the ferroelectric phase appears below $T_{lock}$ temperatures, occurring at 15 K for x = 0.3. The corresponding electric polarization is even 3 times higher than the obtained for x = 0.1 with 5 T applied, increasing 10 fold with the magnetic field.

## 4. Discussion and conclusions

As it has been stressed, a magnetoelectric coupling in this system is confirmed by the anomalous behavior of the temperature dependence of the electric permittivity at the critical temperatures associated with the magnetic phase transitions, by the emergence of an electric polarization in antiferromagnetic phases, and by the effect of magnetic phase transitions on the activation energy of the dipolar relaxation processes observed in these compounds.

From the study of the temperature dependence of the specific heat, the complex electric permittivity, the magnetic response and the thermally stimulated depolarizing currents, the critical temperatures of the $Gd_{1-x}Y_xMnO_3$ system, with $0 \leq x \leq 0.4$, could be determined. In order to complement our interpretation of the experimental results and assign the low temperatures phases, we will use relevant information obtained from single crystals.[8]

The phase transition taking place at $T_N$, well evidenced by anomalies in the temperature dependence of the specific heat and electric permittivity, as well as, from the change of the activation energy of the dipolar relaxation process at around $T_N$, was identified as a transition from the paramagnetic phase to the collinear-sinusoidal incommensurate antiferromagnetic phase. This phase transition is common to all studied compositions, and $T_N$ seems to be almost independent of x.

For the case of $GdMnO_3$ and x = 0.1, the transition from the collinear sinusoidal incommensurate antiferromagnetic to the canted A-type antiferromagnetic phase occurs at $T_{cAFM}$ = 16 K, close to the value reported in the current literature.[18] The results obtained in $GdMnO_3$, yielding a weak ferromagnetic phase, associated with a canted A-type antiferromagnetic arrangement, is in good agreement with published reports.[18] The spin arrangement excludes the existence of a spontaneous ferroelectric phase, according to the inverse Dzyaloshinskii-Moriya model.[22] However, for x = 0 and x = 0.1, a ferroelectric phase emerges for B > 2 T, in good agreement with the current literature.[17,19] In fact, the Y-substitution enhances the stabilization of the ferroelectric phase, which is corroborated by the higher magnitude of the electric polarization at 5 K for x = 0.1 relatively to the one obtained for $GdMnO_3$ at the same temperature.

A different phase sequence occurs for x > 0.1. The weak ferromagnetic character disappears and a pure antiferromagnetic ordering is established. For temperatures below $T_{lock}$, a magnetic phase transition is evidenced by sharp anomalies in the temperature dependence of both real and imaginary parts of the electric permittivity. This phase is also polarizable, wherein the polarization can be enhanced by both applied electric and magnetic fields. According to Ivanov *et al*,[8] this phase is ferroelectric, having an improper character that can be attributed to spin–lattice interaction in the commensurate modulated off-center symmetric magnetic structure. From this statement, we are lead to assign the magnetic phase below $T_{lock}$ to a cycloidal modulated spin arrangement, compatible with ferroelectricity via the inverse Dzyaloshinskii-Moriya interaction. Furthermore, this type of modulation has been also found in the magnetic phases allowing for ferroelectricity in $TbMnO_3$ and $DyMnO_3$,[6] and proposed in the ferroelectric phase of $Eu_{0.6}Y_{0.4}MnO_3$.[15]

For the compositions x = 0.1 and 0.2, the temperature dependence of the real and imaginary parts of the electric permittivity, as well as, the change of the activation energy of the dipolar relaxation process point to structural changes at T* = 22 K. For the compositions x = 0.3 and 0.4, the anomaly of the temperature dependence of the temperature derivative of the magnetization curve is observed at T* = 25 K (x = 0.3) and at 28 K (x = 0.4), along with the change of the activation energy give evidence for structural and magnetic changes in these compounds at T*. Ivanov *et al*[8] have not reported any phase transition at these temperatures. Our data do not allow us to assign, beyond doubt, a phase transition to these anomalies. The existence of anomalies in the temperature dependence of the electric permittivity, within the temperature range of stability of the canted A-type antiferromagnetic phase, have been also reported for $EuMnO_3$ at T' = 23 K.[6] The mechanism associated with such anomalous behavior in the temperature dependence of both the dielectric constant and magnetization can well emerge from a structural change or a spin reorientation at T*. Though, we cannot exclude that this anomaly is associated with the onset of a novel non-polar magnetic transition stable between T* and $T_{lock}$. Further experimental studies are still needed in order to test this hypothesis. In this context, we signalize T* by a dashed line in the proposed (x, T) phase diagram.

Figure 10 depicts the (x, T) phase diagram of the $Gd_{1-x}Y_xMnO_3$ system, with $0 \leq x \leq 0.4$, which summarizes all the aforementioned results. Among the main outcomes of this work, one of them stands out. The mechanisms driving the relevant physical properties of the $Gd_{1-x}Y_xMnO_3$ system could be clearly shown. In fact, tailoring of both magnetic and polar properties of the $Gd_{1-x}Y_xMnO_3$ system, with $0 \leq x \leq 0.4$, at low temperatures can be simply accomplished by handling the octahedral tilting through effective A-site ionic radius. Consequently, the structural changes produced in this way alter effectively the balance between the competitive ferromagnetic and antiferromagnetic exchange interactions leading to the emergence of magnetically induced improper ferroelectric states.[7,15]

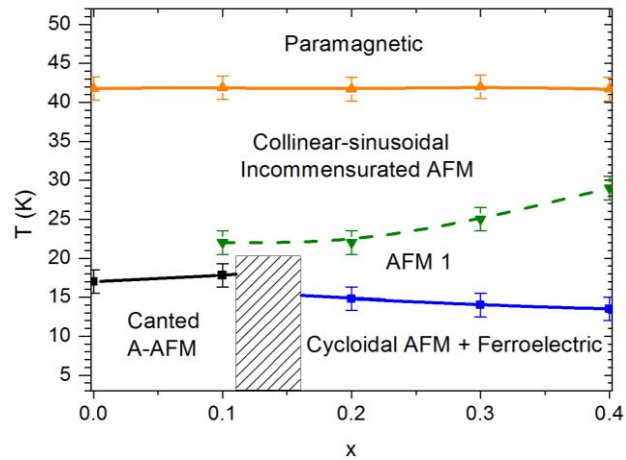

**Figure 9** – Proposed (x, T) phase diagram of the $Gd_{1-x}Y_xMnO_3$ system, with $0 \leq x \leq 0.4$. Dashed line corresponds to uncertainty of phase transition. Grey area corresponds to unknown phase boundaries.

**Acknowledgements**

This work was supported by Fundação para a Ciência e Tecnologia, through the Project PTDC/FIS-NAN/0533/2012 and by QREN, through the Project Norte-070124-FEDER-000070 Nanomateriais Multifuncionais.